\documentclass[]{aa}
\usepackage{amsmath,graphicx,amssymb}
\usepackage{txfonts}
\usepackage{natbib}
\newcommand\hvezda{\object{HD~37776}}
\newcommand{\zav}[1]{\left(#1\right)}
\newcommand{\hzav}[1]{\left[#1\right]}
\newcommand\intvidpo{\!\!\int\limits_{\begin{array}{c}\text{\scriptsize
visible}\\[-2mm]\text{\scriptsize surface}\end{array}}\!\!}

\newlength\staretab

\begin{document}

\title{The light variability of the helium strong star HD~37776 as a result of
its inhomogeneous elemental surface distribution}
\titlerunning{The light variability of the helium strong star HD~37776}

\author{J.  Krti\v{c}ka\inst{1} \and Z. Mikul\'a\v sek\inst{1} \and
        J. Zverko\inst{2} \and J. \v Zi\v z\v novsk\'y\inst{2}}
\authorrunning{J. Krti\v{c}ka et al.}

\offprints{J. Krti\v{c}ka,\\  \email{krticka@physics.muni.cz}}

\institute{Institute of Theoretical Physics and Astrophysics,
           Masaryk University, CZ-611\,37 Brno, Czech Republic
           \and
       Astronomical Institute, Slovak Academy of Sciences,
       SK-059\,60 Tatransk\'{a} Lomnica, Slovak Republic}

\date{Received}

\abstract{We simulate light curves of the helium strong chemically peculiar star
HD~37776 assuming that the observed periodic light variations originate as a
result of inhomogeneous horizontal distribution of chemical elements on the
surface of a rotating star. We show that chemical peculiarity influences the
monochromatic radiative flux, mainly due to bound-free processes. Using the
model of the distribution of silicon and helium on HD~37776 surface, derived
from spectroscopy, we calculate a photometric map of the surface and
consequently the $uvby$~light curves of this star. Basically, the predicted
light curves agree in shape and amplitude with the observed ones. We conclude
that the basic properties of variability of this helium strong chemically
peculiar star can be understood in terms of the  model of spots with peculiar
chemical composition.

}
\maketitle


\section{Introduction}

Periodic light variations are a common feature observed in the magnetic
chemically peculiar (CP) stars. They can be interpreted by the presence of
``photometric" spots causing an uneven distribution of the emergent flux on the
surface of rotating stars. Likely, these hypothetical photometric spots are
somehow related to the ``spectroscopic" spots of a peculiar abundance of various
chemical elements. Within this paradigm, the departures of the energy
distribution in spectra of CP stars from the energy distribution of normal stars
of the same spectral type are as a rule explained as a consequence of line
blanketing by plenty of lines originating in the spectroscopic spots  with
peculiar chemical composition and the flux redistribution induced by them
\citep{peter,molnar}. \citet{lanko} and \citet{mypoprad} suggested that
bound-free transitions may play 
a
more important role in the blanketing,
redistribution of the emergent 
flux,
and the light variability.

However, more effects may influence the spectral energy distribution and
consequently may be invoked to explain the observed light variability. These
effects may be connected with the influence of complex surface magnetic fields.
\citet{ker} and \citet{trasko} showed that the influence of the magnetic
pressure on the stellar atmosphere structure might cause observable light
variations \citep[see also][]{step,carp,lebla}. The effect of the magnetic field
on the atmosphere hydrostatic equilibrium was recently studied by \citet{valy}.
The influence of the Zeeman effect on the line blanketing may modify the
emergent radiative flux \citep{malablaj,malablat}. Consequently, there are many
different mechanisms that might result in the appearance of photometric spots on
the stellar surface.

Contrary to these photometric spots, the location and size (e.g.,
maps of distribution of individual chemical elements on the stellar surface) of
the spectroscopic spots can be determined by Doppler imaging techniques
\citep[e.g.,][]{chochl}. However, it is not clear whether and how the
photometric spots are related to the spectroscopic ones.

Another explanation of the light variability of hot stars with strong surface
magnetic fields exists. \citet{nakaji} and \citet{smigro} proposed that the
observed light variations may be caused by light absorption in rotating
circumstellar clouds. Recently, this model was successfully used by
\citet{towog} for prediction of the light curve of the He-rich chemically
peculiar star \object{$\sigma$~Ori~E}. The properties of these circumstellar
clouds were inferred using the rigidly rotating magnetosphere model
\citep{towo}, in which the material blown from the star by the radiatively
driven wind accumulates in the potential minima along individual magnetic field
lines. \citet{towog} used free parameters characterizing an amount of absorption
in the clouds to predict the light curve of $\sigma$~Ori~E, as it was not known
for how long the accumulation had proceeded. Consequently, they were able to
reproduce the shape of the light curve, but not the amplitude itself. Moreover,
$\sigma$~Ori~E shows hydrogen emission lines that 
indicate
the presence of
circumstellar environment. Last but not least, the light curve of
\object{$\sigma$~Ori~E} is exceptional when compared with the light curves of
other CP stars. However, many CP stars showing light variability are likely too
cool to have sufficiently strong winds enabling formation of circumstellar
clouds. Thus, the light variability of other CP stars may be caused by other
mechanisms.

The fact that CP stars in many cases show not only peculiar chemical
composition, but also 
a variable spectrum
-- indicating the
uneven distribution of chemical elements on their surfaces -- tempts
one to attribute the light variability to this uneven surface
distribution of elements. This uneven distribution, through the line
blanketing and/or the bound-free absorption, determines the emergent
flux depending on the location on the stellar surface and may cause
the variability of a rotating star. However, to our knowledge, a
light curve of a CP star had never been calculated ab initio from
the known distribution of chemical elements as derived by Doppler
imaging techniques. The initial step in this direction was done by
\citet{krivo}, who were able to successfully reproduce the
observed $vby$ light curve of CU~Vir modifying the
temperature gradient in the regions of silicon overabundance.

For our simulations of the light curves of chemically peculiar stars we selected
\object{HD 37776}. This star residing in the reflection/emission nebula
\object{IC 431} \citep{vanden,fink} belongs to well-studied helium-strong CP
stars. It has a very strong complex surface magnetic field with a significant
quadrupolar component \citep{borla,thola}. The observed periodic variations of
helium lines \citep{niss,peto} and lines of some lighter elements \citep{shobro}
are interpreted as a consequence of rotation of a star with uneven elemental
surface distribution. Based on this model, surface distributions and abundances
of various elements were derived from spectroscopy \citep{bola,bupo,choch}.
Consequently, this star seemed to us to be an appealing target for modelling of
the light variations.

In this work we present computed light curves of the He-strong CP star \hvezda.
We used the surface distributions and abundances of various elements derived by
\citet{choch}. As silicon and helium create areas with large overabundances, and
moreover, as it is known that silicon may influence the observed spectral energy
distribution of silicon-rich CP stars \citep{umelec,lanko,mypoprad},  we
concentrated on these two elements in our calculations. The role of silicon may
be more general because this element is overabundant in vast majority of light
variable CP stars \citep[including cool CP stars, e.g.,][]{jednadvacet}.

\section{Basic assumptions}

\subsection{Atmosphere parameters}

Table~\ref{hvezda} lists the atmosphere parameters of \hvezda\ adopted in this
study. We adopted parameters derived by \citet{groka} which are appropriate for
\hvezda\ spectral type \citep{hard}. Even 
though
the value $\log g = 4.5$ used
by \citet{choch} was higher, the value of the surface gravity adopted 
does not have
a
significant effect on the prediction of the light variability (see
Sect.~\ref{disk}). The phases for light curves were computed according to the
linear ephemeris derived by \citet{samadel}, which was used also by
\citet{choch} for surface mapping.

\setlength\staretab{\tabcolsep}
\setlength\tabcolsep{3pt}
\begin{table}[hbt]
\caption{Stellar parameters of \hvezda}
\label{hvezda}
\begin{center}
\begin{tabular}{lccc}
\hline
 ${{T}_\mathrm{eff}}$ & ${22\,000}$\,K & \citet{groka}\\
 ${\log g}$ (CGS) & ${4.0}$ & \citet{groka}\\
Inclination ${i}$ & ${45^\circ}$ & \citet{choch}\\
 Abundance & $-5.4\leq\text{[He/H]}\leq2.3$ & \citet{choch}\\
&$-2.4\leq\text{[Si/H]}\leq2.6$ & \\
$v_{\rm turb}$  & $2\,\text{km}\,\text{s}^{-1}$ &\\
\hline
\end{tabular}
\end{center}
\end{table}
\setlength\tabcolsep{\staretab}

\begin{figure*}
\hfill
\resizebox{0.45\hsize}{!}{\includegraphics{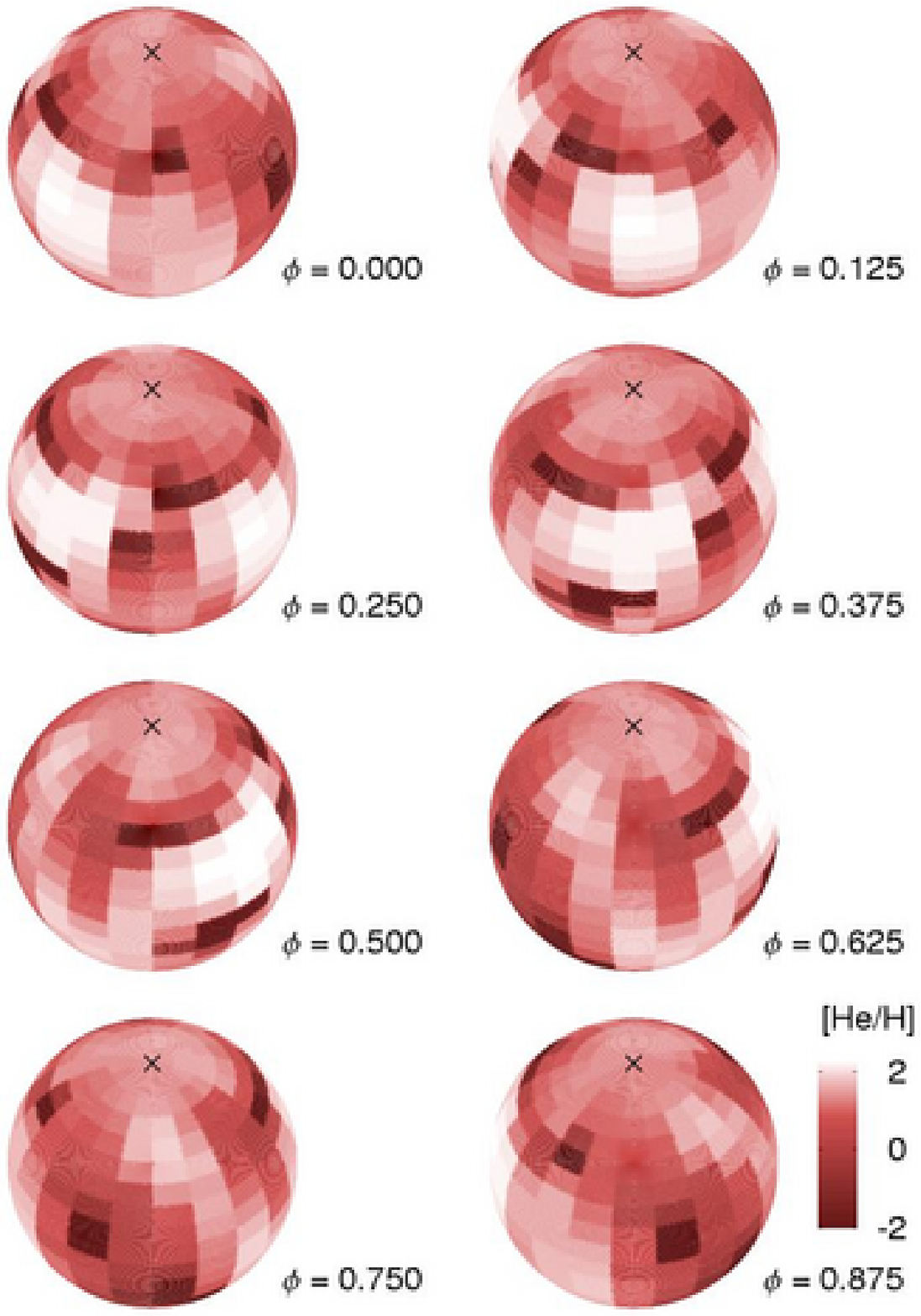}}\hfill
\resizebox{0.45\hsize}{!}{\includegraphics{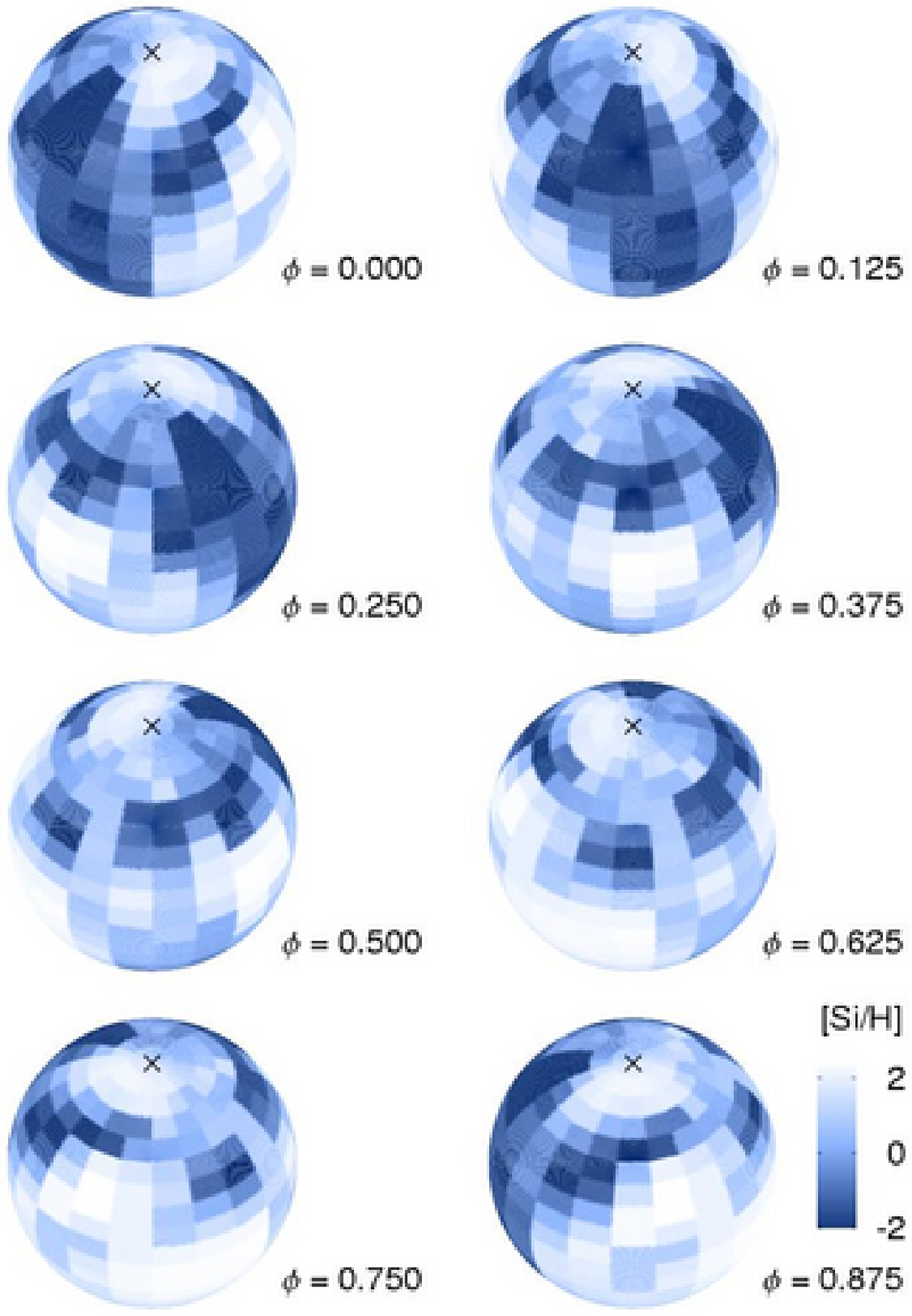}}\hfill
\caption{The variation of the helium (left panel) and silicon abundances on the
visible disc for different rotational phases \citep[after][]{choch}}
\label{povrch}
\end{figure*}

For our study we adopted the surface distribution of He and Si as derived by
\citet[ see Fig.~\ref{povrch}]{choch}. Here the abundance is defined relatively
to the hydrogen, i.e., $[\text{A/H}
]=\log(N_\text{A}/N_\text{H})-\log(N_\text{A}/N_\text{H})_\odot$, where the
arguments of the logarithms are number densities of elements A and H on the
stellar surface, and on the solar surface respectively. Taking into account that
oxygen and iron as derived by \cite{choch} are largely underabundant on 
the entire surface,
we did not consider these elements as sources of the light variability.

\subsection{Model atmospheres and photometric colours}

We use the standard code TLUSTY \citep{tlusty,hublaj,hublad,lahub}
for the model atmosphere calculations. We calculate plane-parallel
model atmospheres in LTE taking into account bound-bound and
bound-free transitions of ions listed in Table~\ref{ionty}. We
included lighter elements only. Although iron and other iron-peak
elements belong to very important sources of opacity in the
atmospheres of hot stars, we neglected their influence because iron
is, according to \citet{choch}, underabundant in the atmosphere of \hvezda\
(see also Sect.~\ref{disk}).
Because it turns out that the bound-free transitions are important for the
variability of \hvezda, we note that the photoionization 
cross-sections
of lighter elements in the TLUSTY ionic models are taken
from the Opacity Project \citep[and references therein]{topt,maslo}.
%
Because here we study mainly the effect of helium and
silicon, for these elements we used the same models as used in the
grid of model atmospheres appropriate to B-type stars calculated by
\citet{bstar2006}. Simpler atomic models were used for other
elements to speed up the calculation of model atmospheres.

For the spectrum synthesis (from which we calculate the photometric colours) we
used SYNSPEC code. We took into account the bound-bound and bound-free
transitions for the same ions as for the model atmosphere calculation (see
Table~\ref{ionty}).

We divided the visible star's surface into $90\times360$ elements in latitude
and longitude, respectively. We selected such a fine grid just from the
numerical reasons to obtain a fairly smooth light curve. The local abundance of
helium and silicon (i.e., their number density relatively to the hydrogen) in
each surface element was set in accordance with the maps of \cite{choch}.  The
emergent flux from these elements was derived by the interpolation from the
model grid. This grid of the model atmospheres contains 80 individual models.
The grid is constructed in such a way that for each of the 10 He 
abundances,
8
different Si abundances were taken according to Table~\ref{hesia}. The abundance
of other elements for simplicity is assumed to be the solar one \citep{sluslo},
i.e., the abundance of elements other than helium and silicon relatively to
hydrogen was kept fixed. Note that the effective temperature and surface gravity
of all these model atmospheres are set to be the same.

\setlength\staretab{\tabcolsep}
\setlength\tabcolsep{4pt}
\begin{table}[t]
\caption{Ions and the number of their excitation states considered for the
calculation of \hvezda\ model atmospheres}
\label{ionty}
\begin{center}
\begin{tabular}{lrlrlrlr}
\hline\hline
Ion & Levels & Ion & Levels  & Ion & Levels & Ion & Levels  \\
\hline
\ion{H}{i}   &   9&\ion{N}{i}   & 21 &\ion{Ne}{i}  & 15 &\ion{Si}{ii}  & 40 \\
\ion{H}{ii}  &  1 &\ion{N}{ii}  & 26 &\ion{Ne}{ii} & 15 &\ion{Si}{iii} & 30 \\
\ion{He}{i}  & 24 &\ion{N}{iii} & 32 &\ion{Ne}{iii}& 14 &\ion{Si}{iv}  & 23 \\
\ion{He}{ii} & 20 &\ion{N}{iv}  &   1&\ion{Ne}{iv} &   1&\ion{Si}{v}   &   1\\
\ion{He}{iii}&   1&\ion{O}{i}   & 12 &\ion{Mg}{i}  & 13 &\ion{S}{ii}  & 14 \\
\ion{C}{i}   & 26 &\ion{O}{ii}  & 13 &\ion{Mg}{ii} & 14 &\ion{S}{iii} & 10 \\
\ion{C}{ii}  & 14 &\ion{O}{iii} & 29 &\ion{Mg}{iii}& 14 &\ion{S}{iv}  & 15 \\
\ion{C}{iii} & 12 &\ion{O}{iv}  &   1&\ion{Mg}{iv} &   1&\ion{S}{v}   &   1\\
\ion{C}{iv} &     1 \\
\hline
\end{tabular}
\end{center}
\end{table}
\setlength\tabcolsep{\staretab}

\begin{table}[t]
\caption{Helium and silicon abundances of the model grid}
\label{hesia}
\begin{tabular}{crrrrrrrrrrr}
\hline
[He/H] & -6 & -5 & -4 & -3 & -2 & -1 & 0 & 1 & 2 & & 3\\\relax
[Si/H] &&&& -3 & -2 & -1 & 0 & 1 & 2 & 2.5 & 3\\
\hline
\end{tabular}
\end{table}

\begin{table}[th]
\caption{Central wavelengths and dispersions of the Gauss filter simulating the
transmissivity functions of $uvby$ (Str\"omgren) photometric system.
These values are taken from \citet{cox}.}
\label{uvby}
\begin{center}
\begin{tabular}{lcccc}
\hline
\hline
colour              & ${u}$& ${v}$& ${b}$& ${y}$\\
\hline
${\lambda_a}$ [\AA] & 3500 & 4100 & 4700 & 5500\\
${\sigma_a}$  [\AA] & 230  & 120  & 120  & 120 \\
\hline
\end{tabular}
\end{center}
\end{table}

We calculate Eddington fluxes $H(\lambda,Y,Z)$ for model atmospheres with helium
and silicon abundances $Y$, $Z$ from the grid ($\lambda$ is the wavelength). The
flux $H_c(Y,Z)$ in a colour $c$ is given by the convolution of $H(\lambda,Y,Z)$
with the transmissivity function of a given filter $c$. The transmissivity
function is approximated by a Gauss function peaked at the central wavelength of
the colour ${\lambda_c}$ with dispersion ${\sigma_c}$ (see Table~\ref{uvby}).
This is performed for each colour $c$ of the $uvby$ system. The radiative
flux in a colour ${c}$ from individual surface elements ${H_c(\Omega)}$ is
obtained by means of interpolation to a proper value of the He and Si abundances
of the surface element concerned. Consequently, the flux ${H_c(\Omega)}$ is a
function of spherical coordinates $\Omega$ on the stellar surface. The total
radiative flux observed at the distance $D$ from the star with radius $R_*$ is
calculated as the integral over all visible surface elements \citep{mih}
\begin{multline}
\label{vyptok}
f_c=\zav{\frac{R_*}{D}}^2\intvidpo I_c(\theta,\Omega)\cos\theta\,\text{d}\Omega
    \\
    =\zav{\frac{R_*}{D}}^2\intvidpo I_c(0,\Omega)u_c(\theta) \cos\theta\,
    \text{d} \Omega\\
    =U_c^{-1}\zav{\frac{R_*}{D}}^2\intvidpo H_c(\Omega)\,u_c(\theta)\cos\theta\,
    \text{d} \Omega,
\end{multline}
where $\theta$ is the angle between normal to the surface element and line of
sight, $u_c(\theta) =I_c(\theta,\Omega)/I_c(0,\Omega)$ describes the limb
darkening (assuming the same limb darkening for each surface element),
$I_c$ is the intensity in colour $c$ and
\begin{equation}
U_c=\frac{1}{4\pi}
\int
u_c(\theta)\cos\theta\,\text{d}\Omega.
\end{equation}

\begin{table}[t]
\caption{Calculated limb darkening coefficients (Eq.~\eqref{okrajterov})
in individual colours}
\label{okrajtetab}
\begin{center}
\begin{tabular}{lcccc}
\hline
\hline
colour & ${u}$ & ${v}$ & ${b}$ & ${y}$\\
\hline
$a_c$ & $0.233$ &  $0.194$ & $0.184$ &  $0.164$ \\
$b_c$ & $0.163$ &  $0.216$ & $0.204$ &  $0.177$ \\
\hline
\end{tabular}
\end{center}
\end{table}

For our calculations we use a quadratic limb-darkening law
\begin{equation}
\label{okrajterov} u_c(\theta)=1-a_c\zav{1-\cos\theta}-b_c\zav{1-\cos\theta}^2.
\end{equation}
We adopted the same limb darkening
regardless the actual abundance of the surface elements.
Namely, the quadratic limb darkening
coefficients $a_c$, $b_c$ in each colour $c$ were derived from the solar
abundance model. Different methods are available for the derivation of limb
darkening coefficients \citep[e.g.][]{grygar}. We selected a method similar to
\citet{negreg}, which conserves the total flux. This leads to a
condition
\begin{equation}
\label{acbc}
b(\lambda)=6-2a(\lambda)-\frac{24H(\lambda)}{I(\lambda,\theta=0)},
\end{equation}
where $H(\lambda)$ is the emergent Eddington flux and $I(\lambda,\theta=0)$ is
the emergent intensity at the $\theta=0$ ray. The value of the coefficient $a$
is derived by the least-square fitting of Eq.~\eqref{okrajterov} with the
condition Eq.~\eqref{acbc}
to values of $u_i(\lambda)$ calculated using SYNSPEC
for different angles $\mu_i$.
For the least-square fitting we used weights equal to angles $\mu_i$.
This leads to a formula
\begin{equation}
\label{amnc}
a=\frac{\sum\mu_i(1-\mu_i)(1-2\mu_i)
   \hzav{\zav{6-\frac{24H}{I(0)}}(1-\mu_i)^2-1+u_i}}
   {\sum\mu_i(1-\mu_i)^2(1-2\mu_i)^2},
\end{equation}
where we, for abbreviation, dropped the wavelength dependence of all quantities.
The limb darkening coefficients in each colour $a_c$, $b_c$ were derived from
Eqs.~\eqref{acbc}, \eqref{amnc} using the gaussian smoothing over the colour $c$
with parameters taken from Table~\ref{uvby}. Bound-bound transitions in the
calculation of $u_c$ were also taken in to account. The calculated linear limb
darkening coefficients are given in Table~\ref{okrajtetab}.

Finally, the observed magnitude difference is
\begin{equation}
\label{velik}
\Delta m_{c}=-2.5\,\log\,\zav{\frac{{f_c}}{f_c^\mathrm{ref}}},
\end{equation}
where $f_c$ is calculated from Eq.~\eqref{vyptok} and
${f_c^\mathrm{ref}}$ is the reference flux obtained from the
condition that the mean magnitude over the phase is zero. As
the spectral energy distribution within the width of a Str\"omgren
filter is affected by the interstellar reddening only marginally,
provided that e.~g.~after \citet{shobro} $E(B-V)=0.081$, we can
neglect the influence of the interstellar reddening. Note that
the interstellar reddening influences mainly the difference between
two colours (the colour index), not the value of the change in a
given colour.

The calculation of magnitudes after Eq.~\eqref{velik} is performed over $36$
rotational phases.

\section{Influence of chemical peculiarity on emergent fluxes}

\begin{figure*}[tp]
{\hfill
\resizebox{0.47\hsize}{!}{\includegraphics{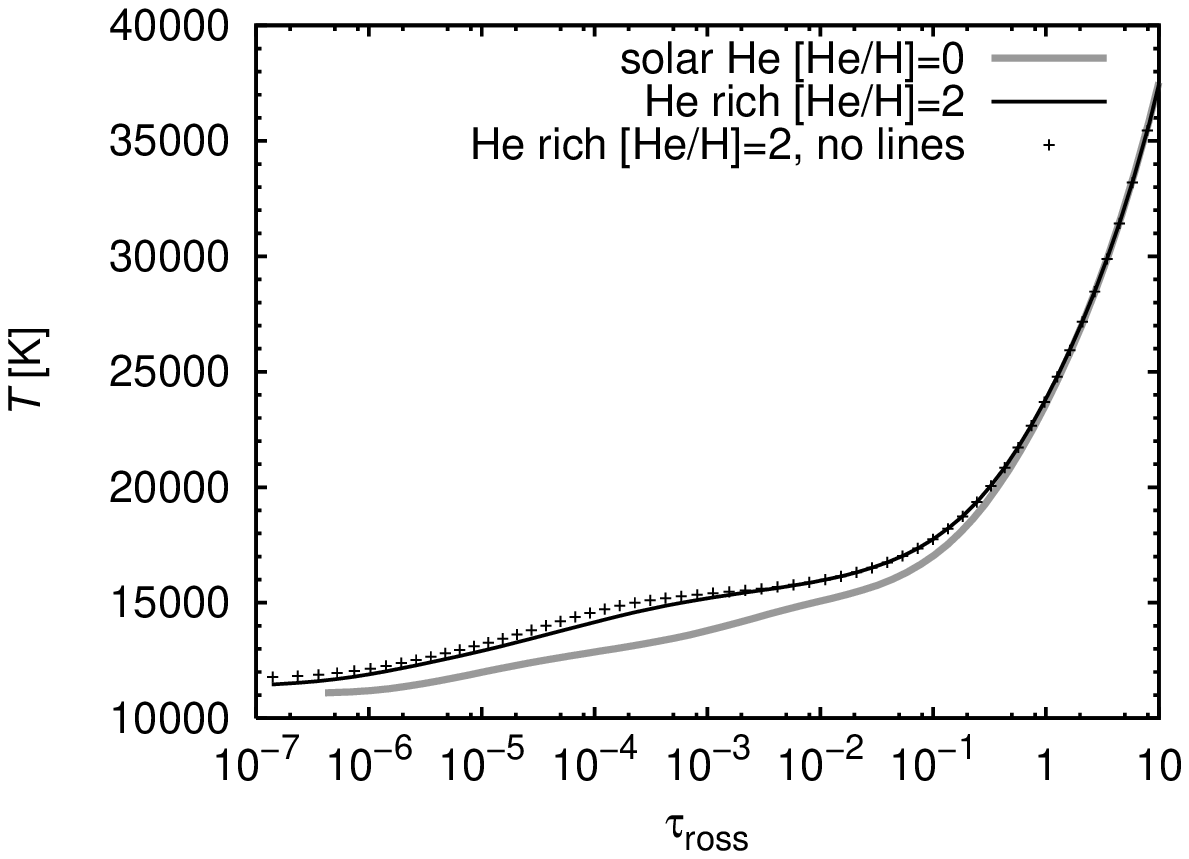}}
\resizebox{0.47\hsize}{!}{\includegraphics{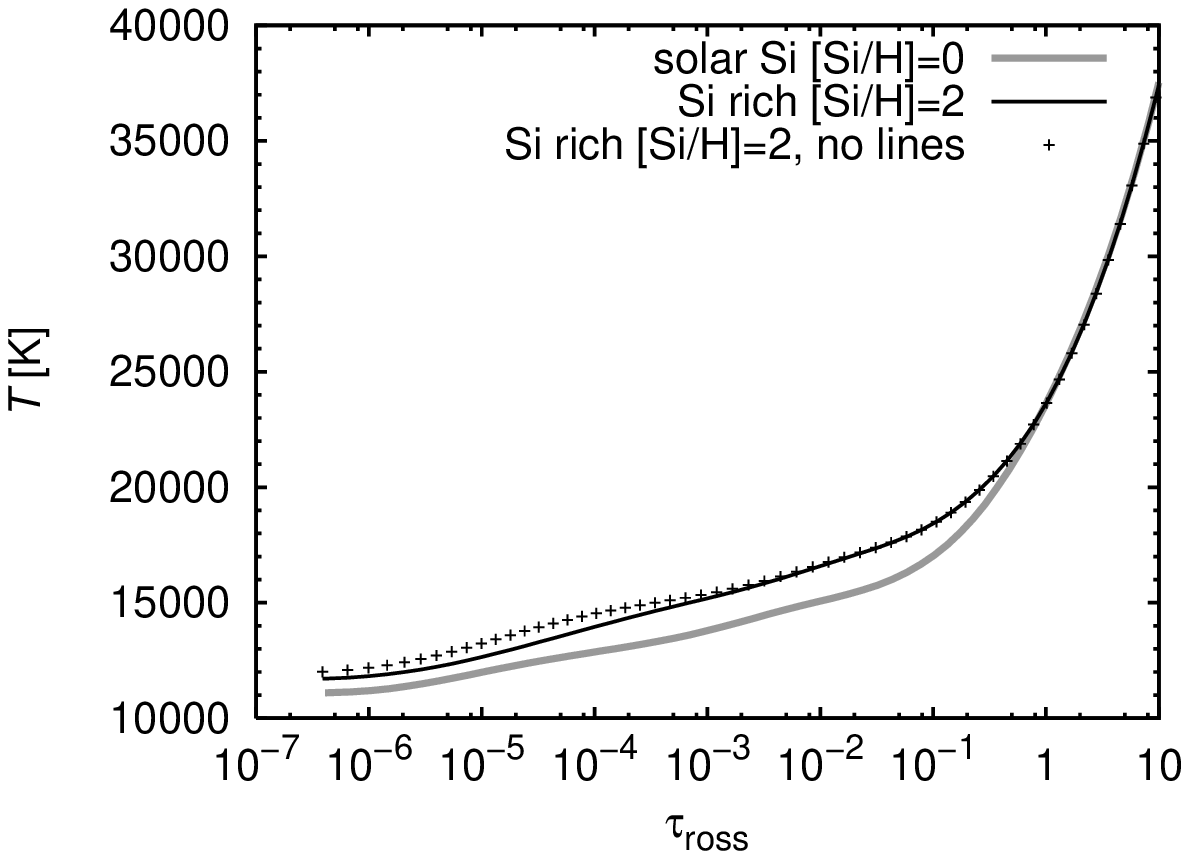}}
\hfill}
\caption{The dependence of the temperature on the Rosseland optical depth
$\tau_\text{ross}$ in the atmospheres with various chemical compositions.
Crosses denote the case of atmosphere with enhanced abundance of either silicon
or helium, however neglecting the opacity due to the line transitions of these
elements.}
\label{tep}
\end{figure*}

\begin{figure*}[tp]
\resizebox{0.5\hsize}{!}{\includegraphics{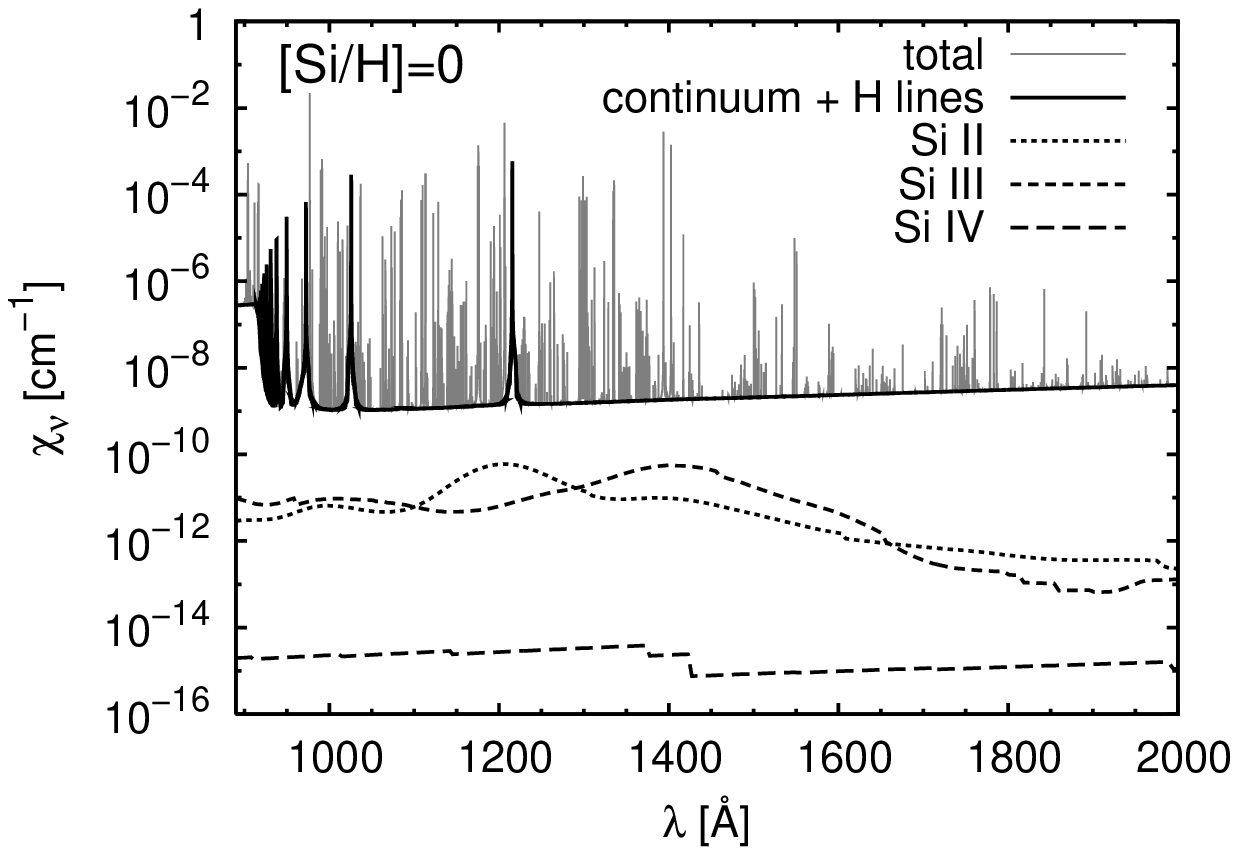}}
\resizebox{0.5\hsize}{!}{\includegraphics{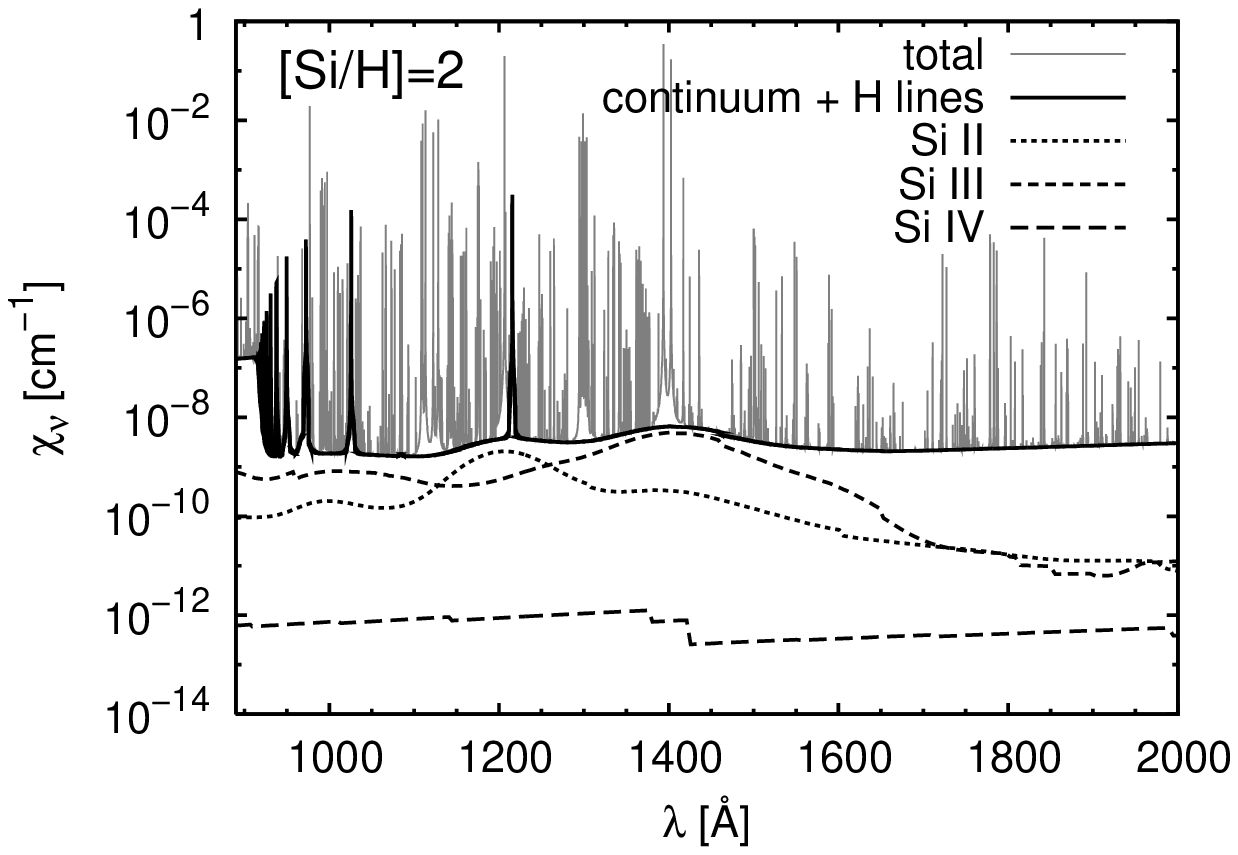}}
\caption{Different sources of the opacity at the point where
$\tau_\text{ross}\approx1$ for the solar chemical composition (left panel) and
for [Si/H]=2. In the silicon overabundant atmosphere the bound-free transitions
due to higher excited levels of \ion{Si}{ii} and \ion{Si}{iii} become important
sources of the opacity in the far UV region of the Balmer continuum.}
\label{opac}
\end{figure*}

As a result of higher opacity in the models with enhanced abundance of
either helium or silicon, the temperature in the outer parts of the atmosphere
increases due to the backwarming. This can be seen in the plot of the dependence
of the temperature on the Rosseland optical depth $\tau_\text{ross}$ for
atmospheres with different chemical compositions in Fig.~\ref{tep}. To identify
the main source of the opacity responsible for the backwarming we also
calculated atmosphere models neglecting the contribution of He-Si lines.
Apparently, the line transitions influence the atmosphere temperature only for
$\tau_\text{ross}\lesssim10^{-3}$, where the stellar atmosphere is basically
transparent in continuum. Consequently, the main influence of the enhanced
abundance of helium and silicon on the atmosphere temperature stratification (in
the region where the photometric flux forms, i.e., for $0.1
\lesssim\tau_\text{ross}\lesssim1$) is due to bound-free transitions. Indeed, as
demonstrated in Fig.~\ref{opac}, in the case of silicon overabundance, the
bound-free transitions from the higher excited levels of \ion{Si}{ii} and
\ion{Si}{iii} significantly contribute to the continuum opacity in the
wavelength region where most of the flux is radiated from the atmosphere of
\hvezda, i.e., in the far UV region of the Balmer continuum.

\begin{figure*}[tp]
{\hfill\resizebox{0.89\hsize}{!}{\includegraphics{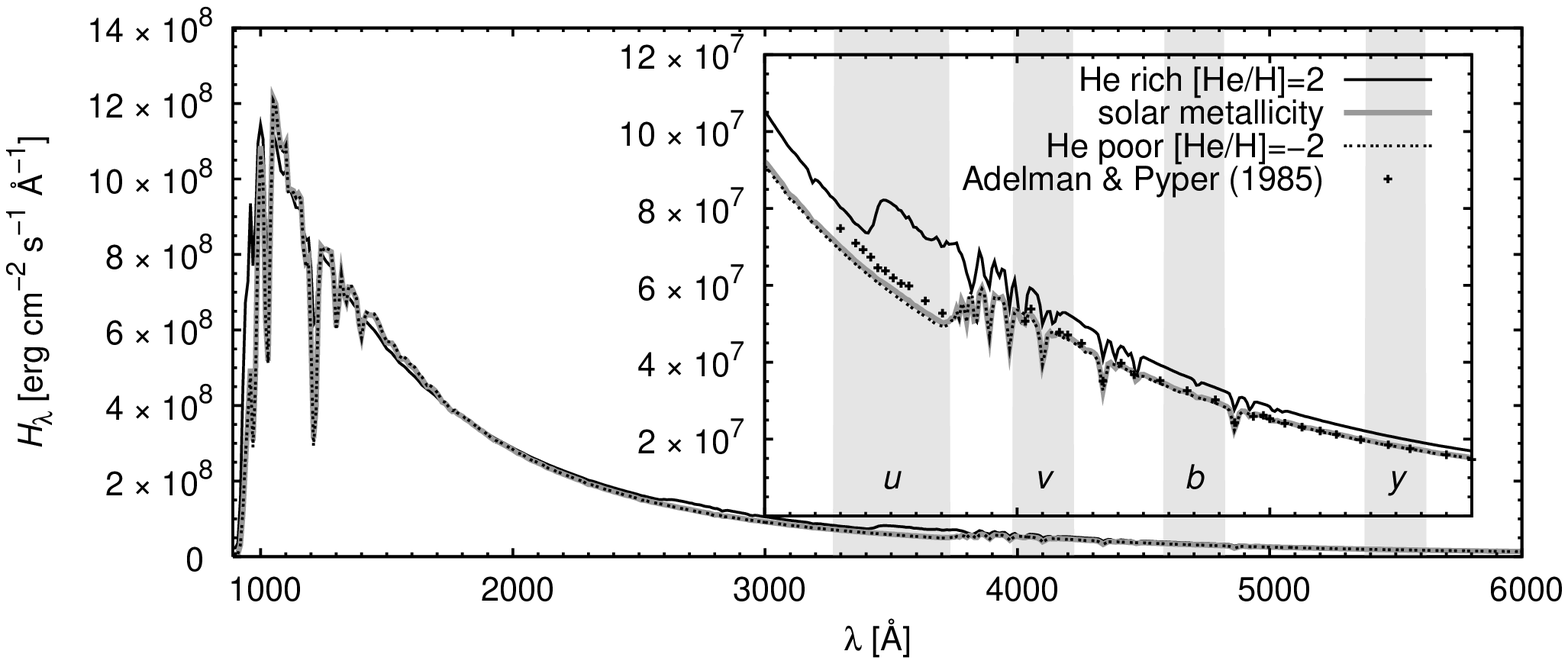}}\hfill}
\caption{The emergent flux from the atmospheres with different helium
abundances. The flux was smoothed by a Gaussian filter with a dispersion of
$10\,\AA$ to show the changes in continuum, which are important for photometric
variability. The passbands of the $uvby$ photometric system are also shown on
the graph (gray area). The photometric region was also enlarged and overplotted.
For a reference we show also the observed mean flux derived by \citet{adelpy},
which we normalize using the predicted solar-abundance flux at 5000\,\AA.}
\label{hetoky}
\end{figure*}

\begin{figure*}[tp]
{\hfill\resizebox{0.89\hsize}{!}{\includegraphics{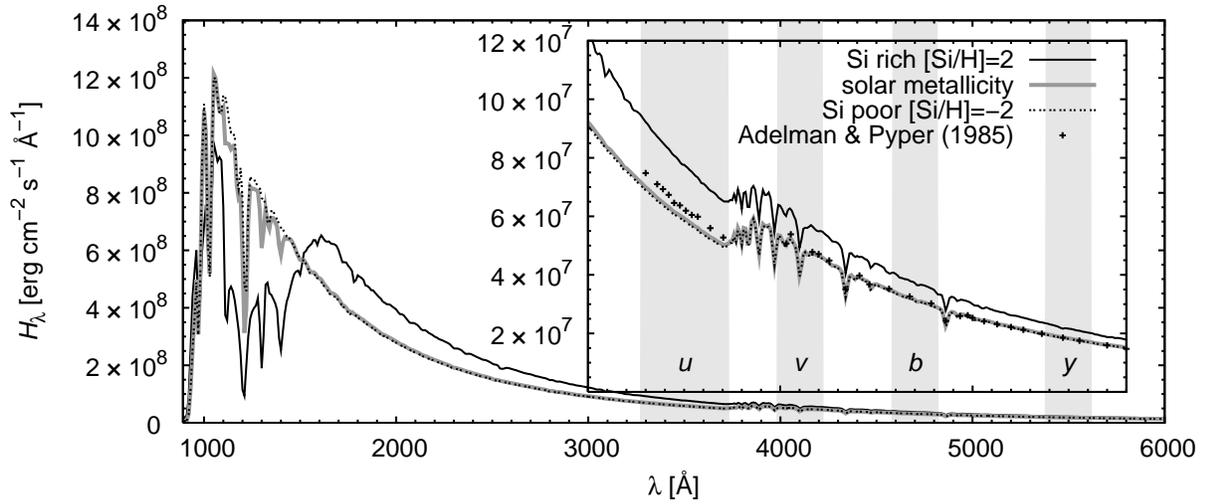}}\hfill}
\caption{The same as Fig.~\ref{hetoky}, for silicon.}
\label{sitoky}
\end{figure*}

\begin{figure*}[tp]
\resizebox{0.5\hsize}{!}{\includegraphics{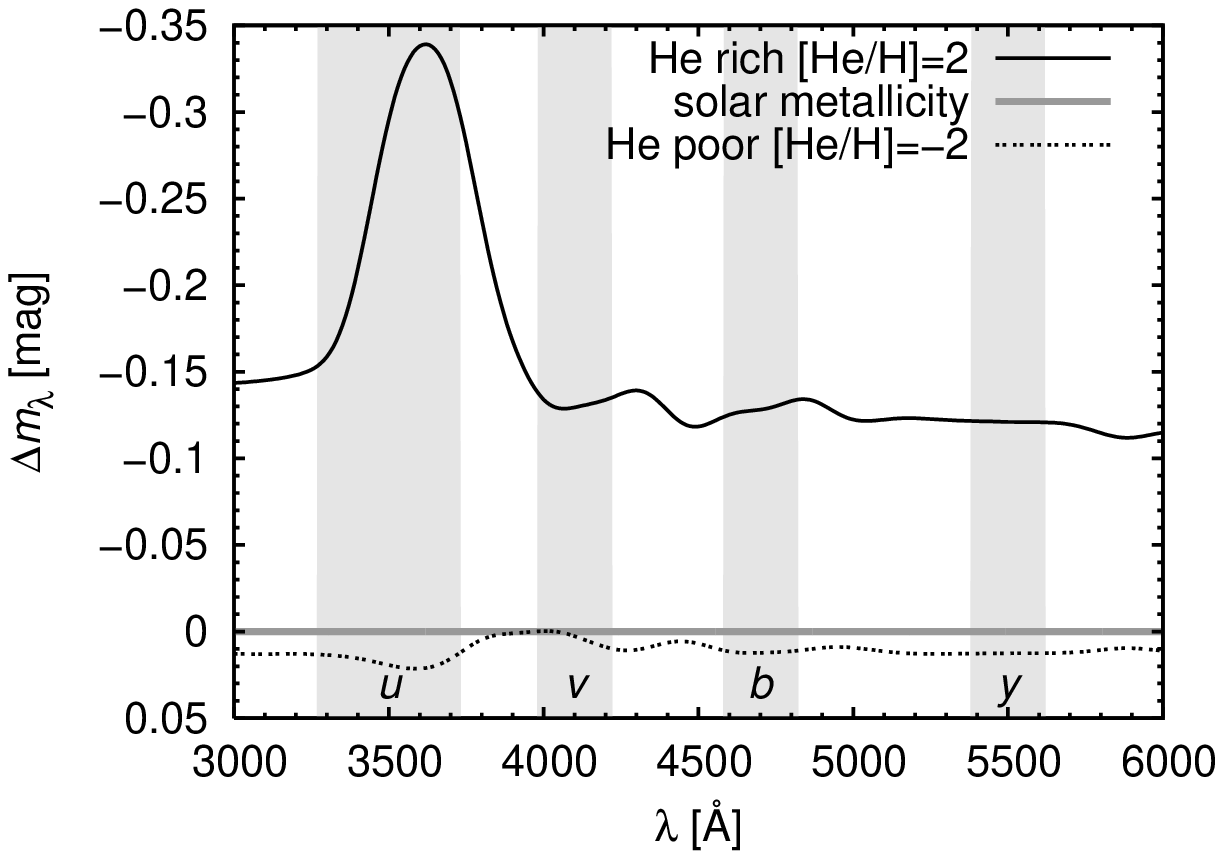}}
\resizebox{0.5\hsize}{!}{\includegraphics{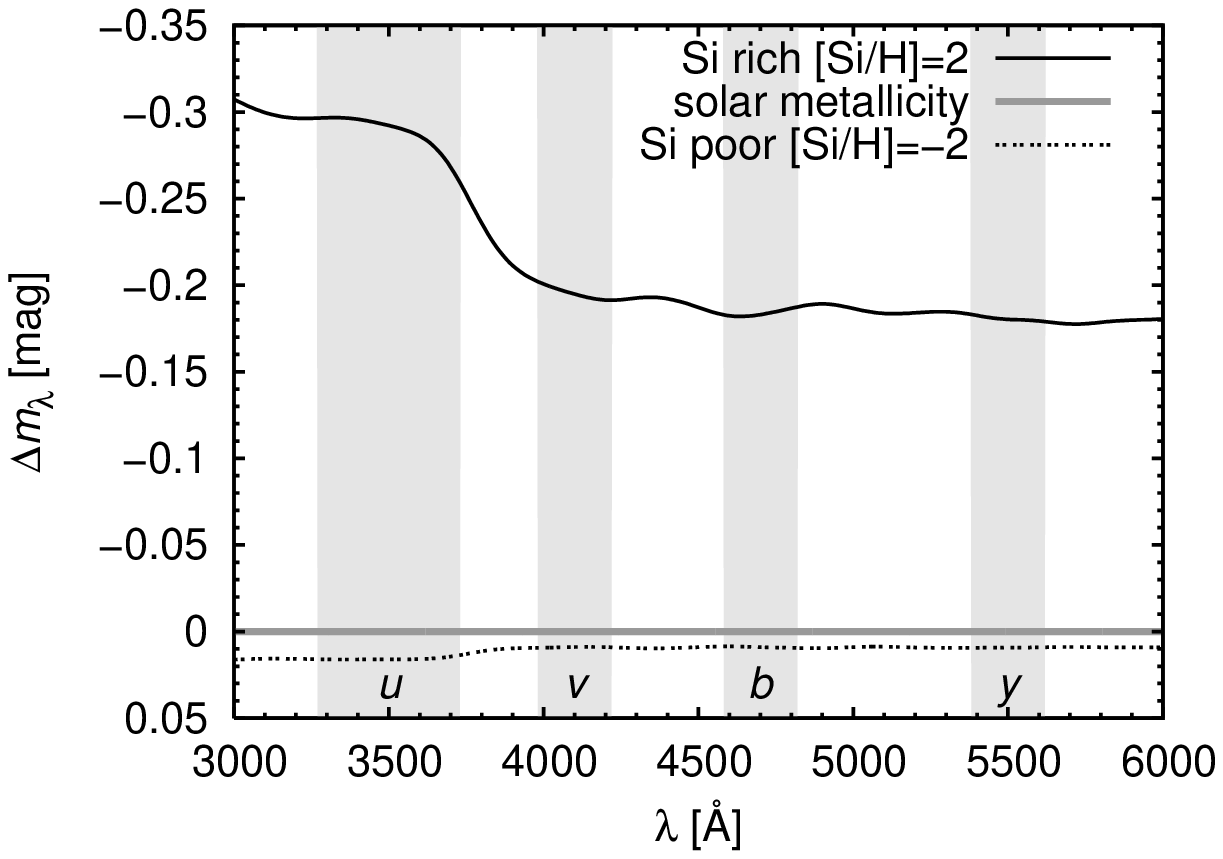}}
\caption{The magnitude difference between the flux at the given abundance of
helium ({\em left panel}) and silicon ({\em right  panel}) $H(\lambda,Y,Z)$ and
the flux for the solar abundance $H(\lambda,Y_\odot,Z_\odot)$ calculated  as
$\Delta m_\lambda=-2.5\log( H(\lambda,Y,Z)/H(\lambda,Y_\odot,Z_\odot))$. The
passbands of the $uvby$ photometric system are also shown on the graph (gray
area). The flux was smoothed by a Gaussian filter with dispersion
120\,\AA~to show the changes in the wavelength range corresponding to the
pass-bands of the $u$, $v$, $b$ and $y$ filters.}
\label{logtoky}
\end{figure*}

The radiative flux emerging from the model atmospheres depends on the silicon
and helium abundances, as can be seen in Figs.~\ref{hetoky}, \ref{sitoky}.
For solar chemical composition, the Balmer jump is visible in the plot. With
increasing silicon abundance, the opacity in the ultraviolet (UV) region below
roughly $1600\,$\AA\ increases mainly due to bound-free silicon
transitions (and partly also due to the bound-bound transitions, see
Fig.~\ref{opac}). The flux from this UV region is redistributed towards longer
wavelengths, partly into the optical region. Consequently, the silicon
overabundant regions are brighter than the silicon poor ones in the $uvby$
bands. A very similar situation 
occurs
also 
in
the case of helium overabundance.
However, since for [He/H]=$2$ the helium dominates in the stellar atmosphere,
jumps due to ionization of neutral helium occur. Helium overabundant regions are
also brighter in the $uvby$ bands than those with lower helium abundance.
Decreased helium and silicon abundances below the solar values outside the
bright spots 
do
not significantly influence the flux in the $uvby$ colours. As
a result, we expect that 
only bright spots
occur on the stellar surface, not the
dark ones that may seemingly originate only due to a contrast effect.

There is a relatively good agreement between the shapes of the calculated
and predicted spectral energy distribution (for solar abundance) in the visible
region \citep[taken from][note that from the observations only fluxes expressed
relatively to the flux at $5000\,$\AA\ are available]{adelpy}. The observed
fluxes are slightly higher than the predicted ones in the Lyman continuum just
below the Balmer jump. This difference might reflect the influence of the
chemical peculiarity.

To identify the main sources of the light variability, we
compared the fluxes calculated using models with neglected line
transitions with that ones calculated with inclusion of the line
transitions. This comparison showed that the line transitions
contribute to the light amplitude only by up to percents.

The change of the fluxes manifests itself through the change of the apparent
magnitude, Fig.~\ref{logtoky}. The change is nearly the same in the $b$ and $y$
colours, which is caused by the fact that these colours lie on the Wien part of
the energy distribution function and no important ionization edges are located
there. The different change in the $u$ colour is connected with the behaviour of
the Balmer jump and the occurrence of the jump due to helium ionization. The
change in the $v$ colour may be also slightly influenced by the presence of the
Balmer jump.

\begin{figure*}[tp]
\resizebox{0.5\hsize}{!}{\includegraphics{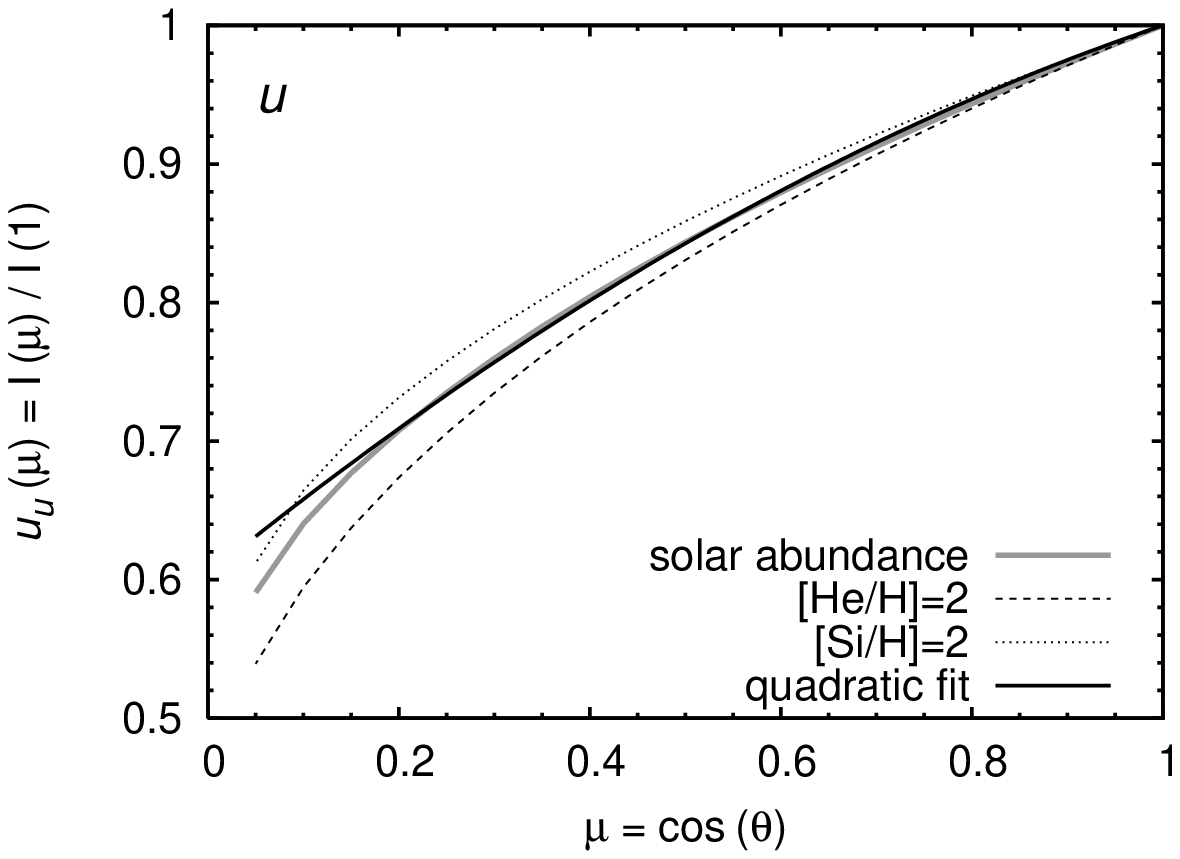}}
\resizebox{0.5\hsize}{!}{\includegraphics{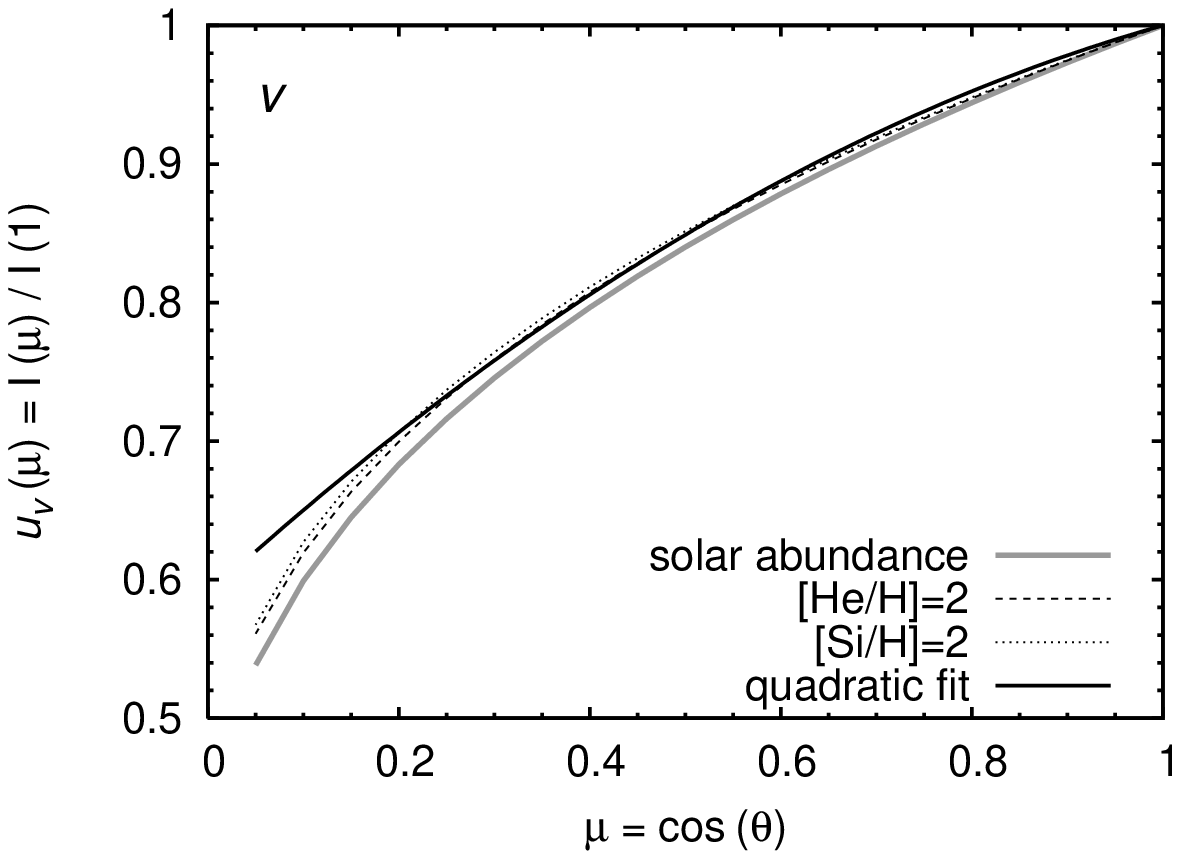}}
\resizebox{0.5\hsize}{!}{\includegraphics{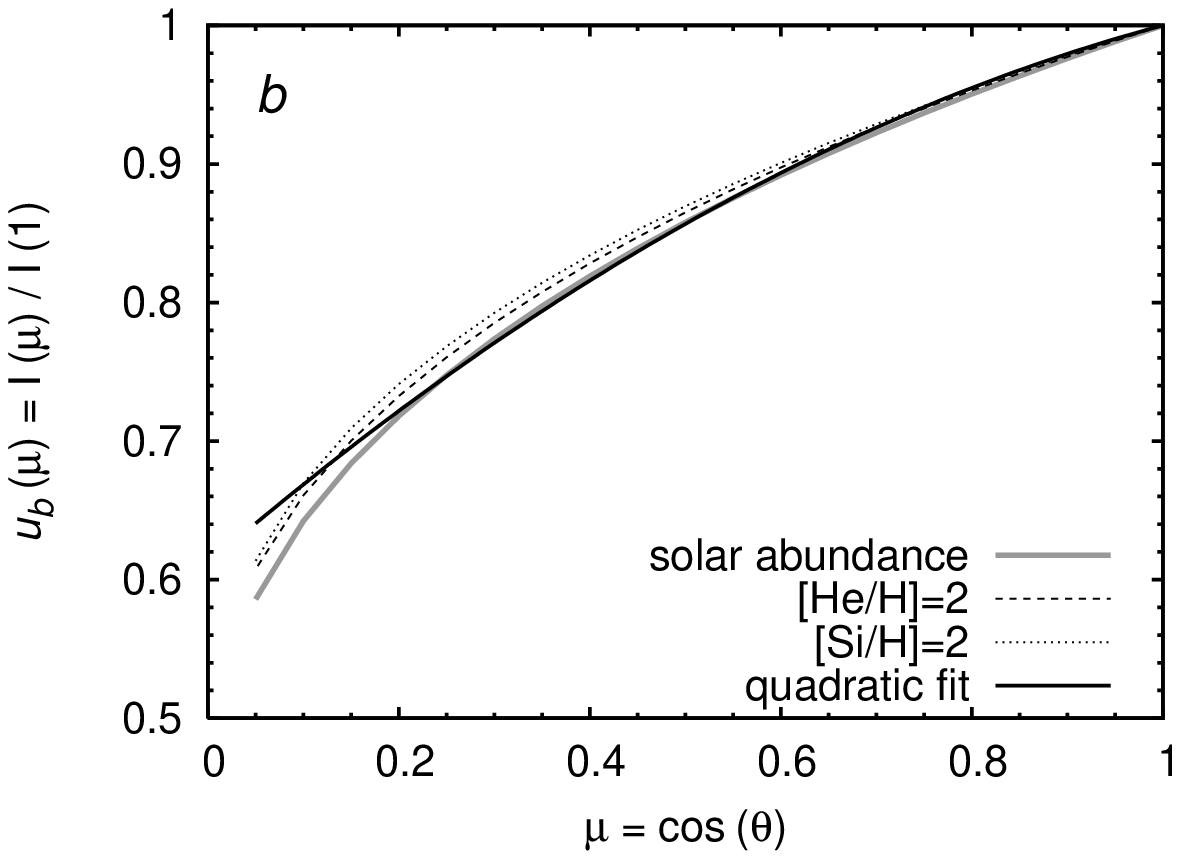}}
\resizebox{0.5\hsize}{!}{\includegraphics{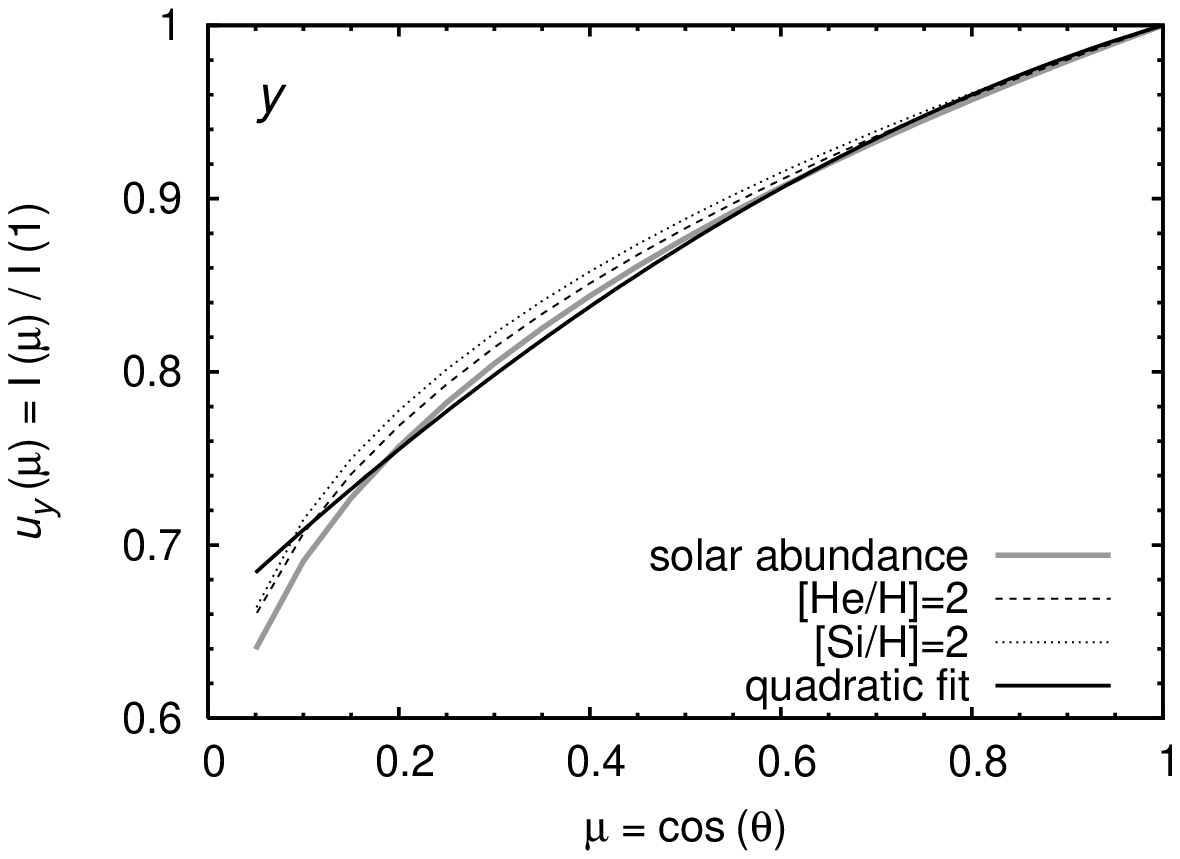}}
\caption{The {\em continuum} limb darkening at the wavelength corresponding to the
center of the individual filter (Tab.~\ref{uvby}) in the model atmospheres with
different chemical composition. The solid black line denotes the quadratic limb
darkening (see Eq.~\eqref{okrajterov}, taking into account also
lines) adopted in this paper.}
\label{okrajte}
\end{figure*}

The change of the chemical composition of the atmosphere 
also influences
the limb darkening as shown in Fig.~\ref{okrajte}. However, near the disk center
for $\cos\theta>0.6$ the variations of the limb darkening with abundance are not
so significant. Consequently, we did not include these changes into our
modelling, because the approximation of the limb darkening does not have a
significant influence on the calculated light curves. For example, out test
showed that there is only a relatively small difference, up to $\approx 0.001$
mag, between the light curves calculated using quadratic and linear limb
darkening.

\section{Simulation of the light curve}

\subsection{Helium}

\begin{figure}[t]
\resizebox{\hsize}{!}{\includegraphics{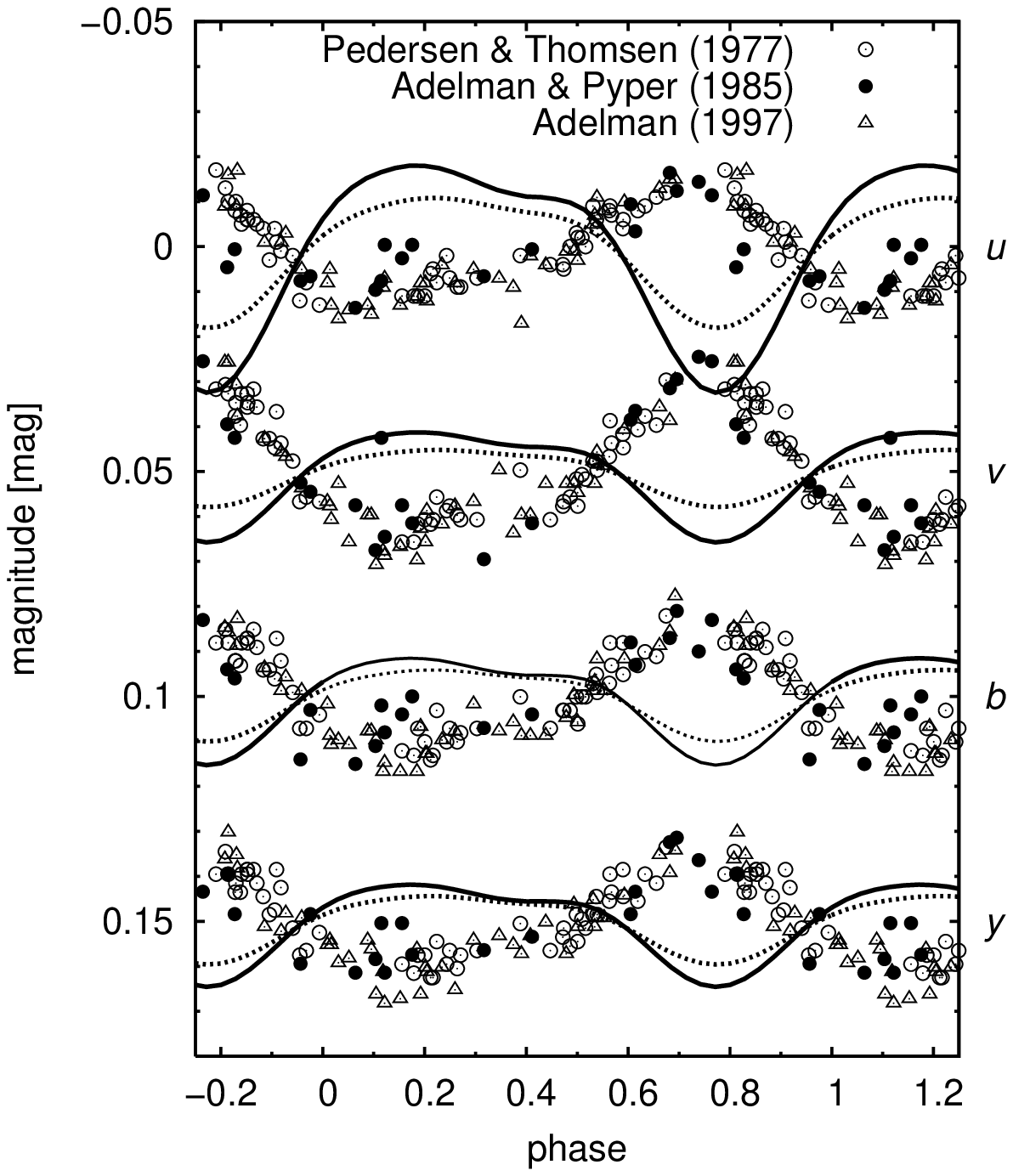}}
\caption{The simulated light variations of \hvezda\ in different colours derived
taking into account only the uneven distribution of helium on its surface
(solid line) compared with individual $uvby$ observations
\citep{peto,adelpy,samadel}. The dotted line denotes the simulated light
curve derived taking into account the helium abundance reduced by $1\,$dex. All
observed magnitudes are shifted in such a way that their mean value is zero. A
vertical offset of 0.05\,mag was applied between each consecutive filter.}
\label{hehvvel}
\end{figure}

Since helium is the dominant element in the majority of regions on the
\hvezda\ surface \citep{choch}, we started  studying the light variations caused
by this element only. The corresponding light curves calculated taking into
account the uneven distribution of helium on the \hvezda\ surface (and the solar
chemical composition of other elements) are given on Fig.~\ref{hehvvel}. Because
the most helium-rich regions appear on the visible disk during the light minimum
of the observed light variations, and because the helium overabundance induces
the brightening of the stellar surface (Fig.~\ref{hetoky}), the predicted light
variations caused by helium are practically in antiphase with the observed ones.
Consequently, although helium significantly contributes to the light variability
of \hvezda, there has to 
also be
another mechanism that dominates the optical variability of this star.

To check the response of the light curve to the various values of the abundance
of helium we reduced it by 1 dex relative to the value derived by \citet{choch}.
This, however, did not induce a remarkable change, as seen in
Fig.~\ref{hehvvel}. This is due to the fact that even with [He/H]=1 the
atmosphere starts to be dominated by helium already.

\subsection{Silicon}

\begin{figure}[t]
\resizebox{\hsize}{!}{\includegraphics{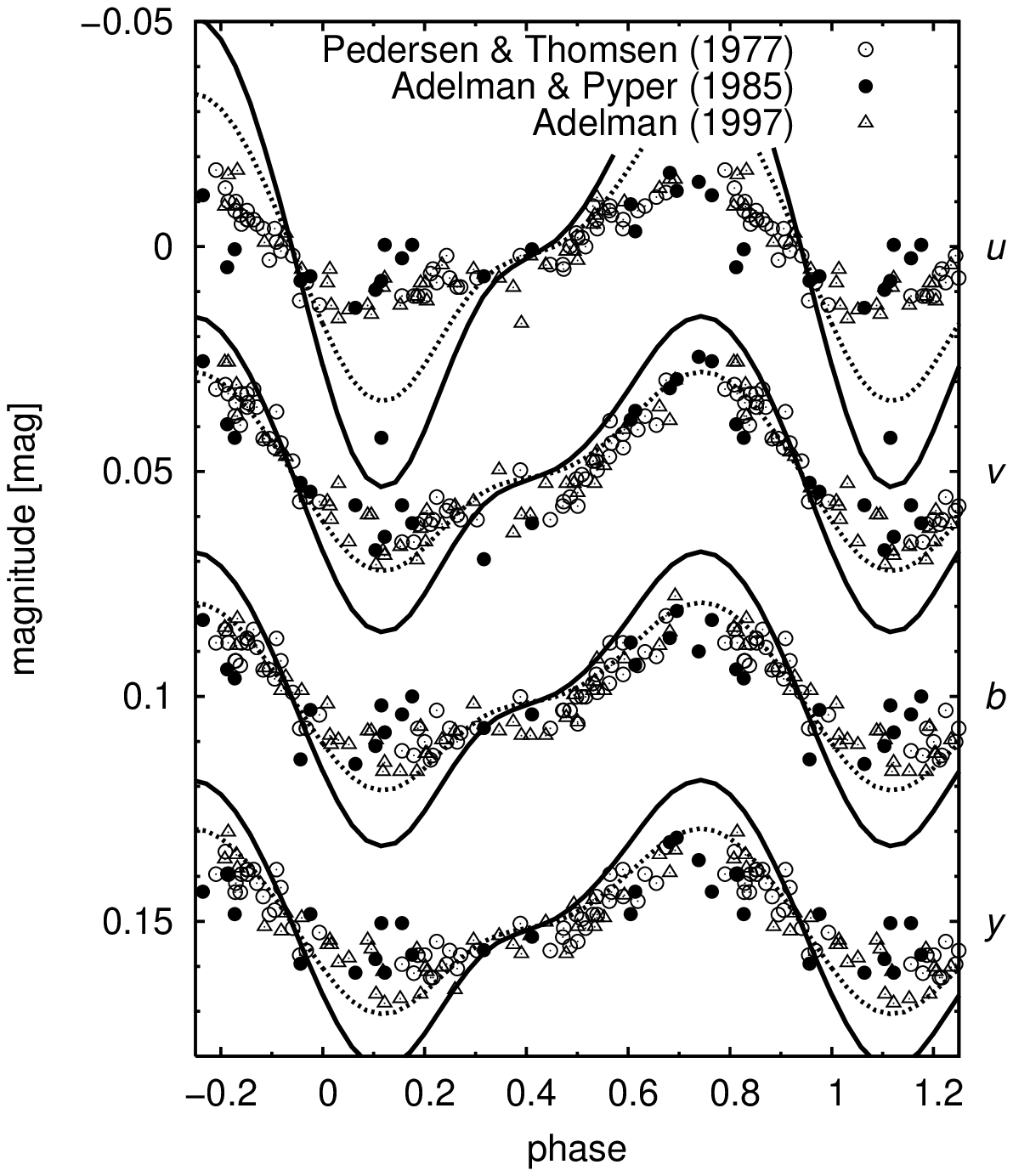}}
\caption{The same as Fig.~\ref{hehvvel}, taking into account only the uneven
distribution of silicon. The dotted line denotes the light curve calculated
for the surface distribution of silicon derived from the observations by
\cite{choch} decreased by $0.5\,$dex. A vertical offset of 0.05\,mag was applied
between each consecutive filter.}
\label{sihvvel}
\end{figure}

The fact that silicon is overabundant on the observed disc around the phase
$\phi=0.75$, when the star has its light maximum \citep[see also
Fig.~\ref{sihvvel} and Fig.~\ref{povrch}]{samadel}, 
along with
our finding that silicon rich regions are bright, 
lead us to
deduce that silicon may be
the main cause of the \hvezda\ variability. To test this, we calculated the
light curve of \hvezda\ taking into account only the uneven distribution of
silicon \citep{choch}. The resulting light curve is displayed in
Fig.~\ref{sihvvel}. Apparently, taking into account only the uneven surface
distribution of silicon, the amplitude of the light curve is too large when
compared with the observed one (especially in the $u$ colour). However, there is
a good agreement between the light curve shapes. Note also that the amplitude is
relatively sensitive to the value of [Si/H].

\subsection{Helium and silicon}

\begin{figure}
\resizebox{\hsize}{!}{\includegraphics{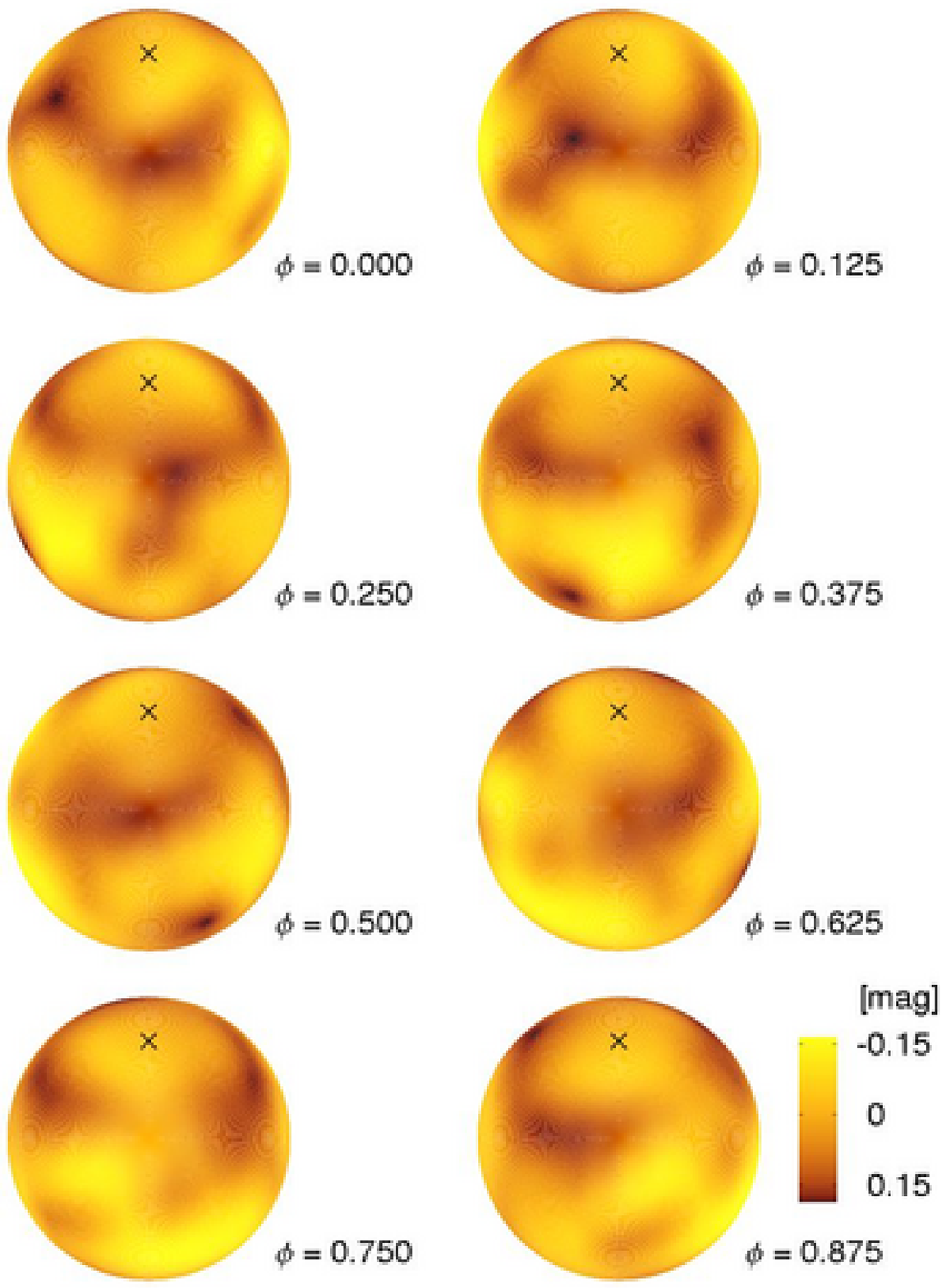}}
\caption{The location of the calculated photometric spots in the $u$ colour. We plot the
smoothed
flux emergent from the individual surface elements calculated using the observed
surface distribution of helium and silicon derived by \citet{choch} in
magnitudes. Here we do not include the limb darkening.}
\label{vidpovrch}
\end{figure}

\begin{figure}
\resizebox{\hsize}{!}{\includegraphics{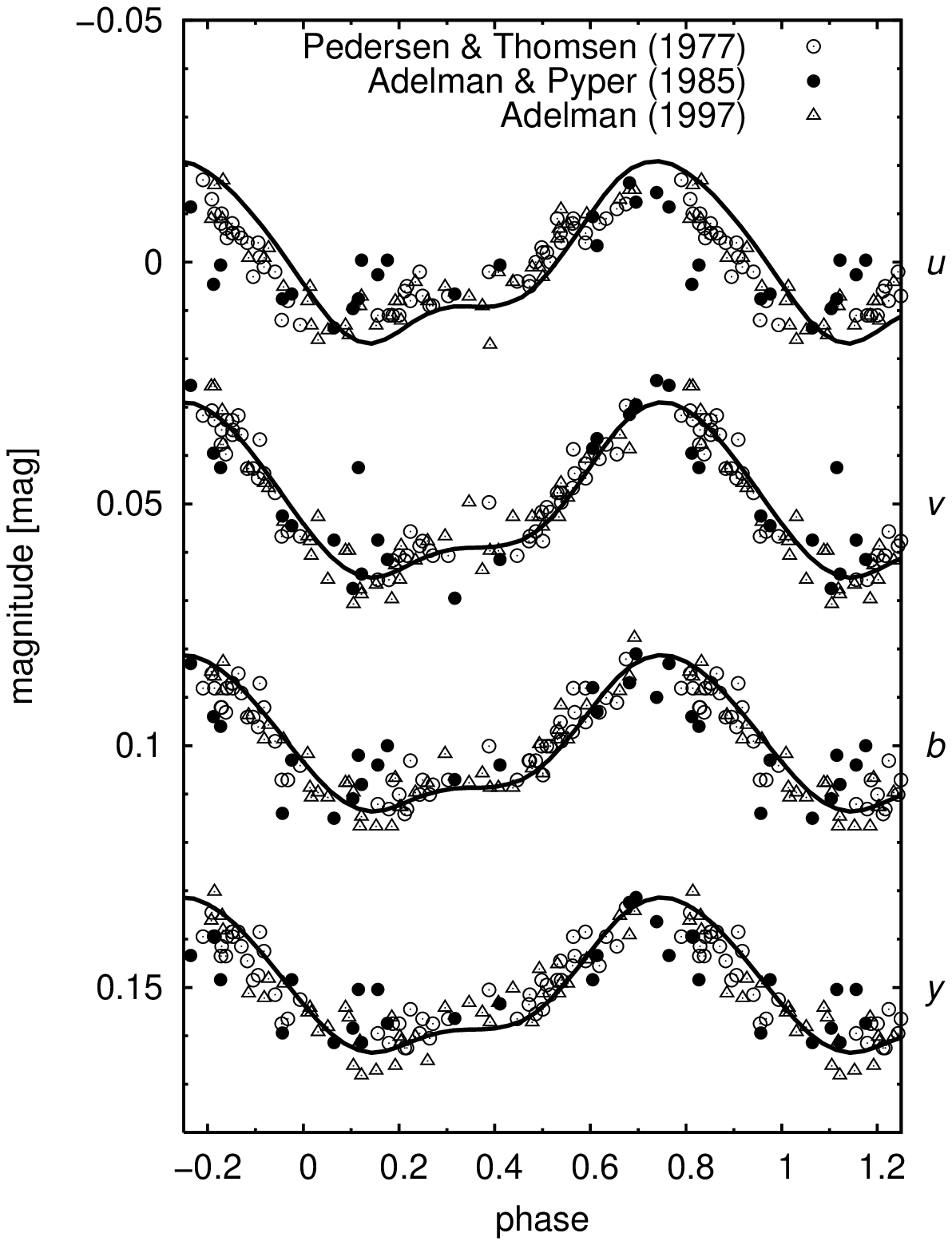}}
\caption{The same as Fig.~\ref{hehvvel}, taking into account the uneven
surface distribution of both the helium and silicon (solid line).
A vertical offset of 0.05\,mag was applied between each consecutive
filter.}
\label{hvvel}
\end{figure}

The promising correspondence between the observed and predicted
light curves derived above 
encourages 
us to include both
silicon and helium 
in
the calculation of the light curve. To
visualize the origin of the light variations, we plot the
photometric spots on the surface of \hvezda\ in different rotational
phases in Fig.~\ref{vidpovrch} in the $u$ colour. As it follows from
this plot the shape of the light curve significantly depends on the
inclination~$i$.

Inspecting Figs.~\ref{logtoky},~\ref{hehvvel} we can conclude that the
variations due to helium and silicon are basically in the antiphase.
Consequently, helium may suppress the too large amplitude derived assuming
silicon surface variations only  (especially in the $u$ colour see
Fig.~\ref{sihvvel}). Indeed, the computed light curve shown in
Fig.~\ref{hvvel}, where both silicon and helium abundance variations are taken
into account, confirms these considerations. There is a very good agreement
between the predicted and observed light curves in the shape and amplitude.
There is just a relatively small disagreement in the $u$ colour. Some part of
the remaining discrepancy between the predicted and observed curves may be due
to uncertainties of spots location as derived from observations, as the observed
maps are tabulated at relatively coarse interval of $18^\circ$ in longitude.

Consequently, we conclude that the observed light variability of \hvezda\ can be
successfully simulated assuming the uneven surface distribution of
individual elements, mostly silicon and helium.

\section{Discussion}
\label{disk}

\paragraph{The role of other elements}

An uneven surface distribution of elements other than silicon and helium may
influence the light curve. Whereas the influence of oxygen and iron, which
are underabundant on the \hvezda\ surface \citep{choch}, is likely to be
negligible, other elements \citep[especially carbon for which the spectral line
variability is observed, see][]{shobro} may modify the predicted light
curve. On the other hand, if the iron abundance is much higher than
what we assumed,
then the iron opacity could be important for the
spectral energy distribution and variability (provided the uneven surface
distribution of iron).
We note that \citet{choch} derived from Doppler-Zeeman mapping that iron is
relatively most abundant in spots where its deficit reaches up to 2 dex, while
in the rest of the atmosphere the deficit is 4 -- 5 dex relative to the solar
value. These values are surprisingly low and deserve further investigation using model
atmospheres with magnetic 
fields
included.

\paragraph{NLTE effects}

The NLTE effects have a significant influence on the spectra of hot stars.
These effects influence mainly the state of the outermost parts of the stellar
atmospheres, where the lines are formed, and which is optically thin in the
visible continuum. Since \citet{choch} used LTE models for mapping of the
\hvezda\ surface (and, consequently, the real abundances may be different), we
use also the LTE models here. The neglect of the NLTE effects for both the
abundance determination and synthetic spectrum calculation may be one of the
reasons of the remaining small discrepancy in the $u$ colour.

\paragraph{Influence of the magnetic field}

Staude (1972) and Trasco (1972) showed that 
a
strong surface magnetic field may
influence the emergent spectral energy distribution which may in photometry
amount to hundredths of magnitude. Khan \& Shulyak (2006) and Kochukhov et al.
(2005) studied the influence of the polarized radiative transfer and magnetic
line blanketing on the energy distribution, atmospheric structure and
photometric colours and showed the magnetic field has a clear relation to the
visual flux depression. \citet{step}
deduced that the effect of the magnetic
field on the stellar atmosphere leads to variation of $T_{\rm eff}$  and $\log
g$ over the stellar surface resulting in light variations. Consequently, the
surface variations of
the
magnetic field may also contribute to the light
variability.

\paragraph{Vertical distribution of elements}

The vertically dependent abundance variations in chemically peculiar stars are
suggested by many authors \citep[e.g.,][]{rypop}. This effect may also
influence our results, because the effective depth at which the lines form
(i.e., at which the abundance map is in fact derived) differs from that at
which continuum forms (i.e., at which the light variations are formed).
Moreover, the chemical stratification itself may influence the emergent
continuum flux. However, to our knowledge, no evidence of the vertical chemical
stratification is available for \hvezda\ atmosphere at the present time.

\paragraph{Uncertainties of stellar parameters}

Uncertainties of the determination of the stellar parameters may influence the
comparison between observed and predicted quantities. However, since we study
mainly differential flux variations in individual colours, the influence of the
determination of, e.g., the surface gravity acceleration or the stellar
effective temperature are likely to be less important. Indeed, the comparison of
the the magnitude differences calculated for [Si/H]=2 after Eq.~\ref{velik} for
stellar model calculated with the effective temperature by $1000\,$K lower or
the surface gravity by $0.5\,$dex higher shows that the magnitude differences
vary by about $0.02\,$mag (compare with the change of $0.1-0.35\,$mag showed in
Fig.~\ref{logtoky}). The magnitude differences are even less sensitive to the
changes of the turbulent velocity, because the increase of the turbulent
velocity by $2\,\text{km}\,\text{s}^{-1}$ causes the variation of the magnitude
difference by only about $0.001\,$mag.

\paragraph{Alternative surface chemical composition maps}

Because with the helium-only overabundance model we came 
up with
a big discrepancy
between the computed and observed light curves (Fig.~\ref{hehvvel}), we used
another available model derived for the surface chemical composition of helium
only, by \citet{bupo}. However, neither with this different surface pattern we
were able to reproduce the observed light curve (cf.~Fig.~\ref{hehvvel}). A
further map of the surface chemical composition which can be used for the
reliable simulation of the light curve is that of \citet{bola}. To test the
influence of independent surface chemical composition map on our result, we used
the original map derived by \citet{choch}, however the maximum and minimum
abundances of helium and silicon were taken from \citet{bola}. Generally, as a
result of lower silicon abundance in the spot \citep[compared with that by][]
{choch}, the amplitude due to silicon is lower when using \citet{bola} silicon
abundance. However, this is compensated by lower maximum helium abundance.
Consequently, our tests showed that also with this different surface map we are
able to obtain a fair agreement with observations.

\paragraph{Light variations in UV}

If the proposed mechanism of the light variability is correct, the studied star
should be variable also in the UV region and the observed light curve in the far
UV region should be in antiphase with the optical and
near-UV light curve. Indeed, a similar behaviour was found for other CP
stars \object{CU Vir} and \object{56 Ari} by \citet{sokolj,sokold}.

\paragraph{Comparison with observed flux distribution}

The next step in the comparison of the observed and predicted characteristics
would be to compare the mean observed average flux derived by \citet{adelpy} and
the theoretical one. Whereas there is a good agreement between these fluxes in
the visible region, there is a slight disagreement (less than $10\%$) between
them in the region between helium and hydrogen ionization edges in the $u$
colour. The enhancement of the mean predicted flux in this region is caused by
the enhanced helium abundance. Consequently, this might indicate that the
maximum helium abundance on \hvezda\ surface is lower than that reported by
\cite{choch}. We also note that, for example, the effective temperature of
\hvezda\ was obtained using atmosphere models with solar chemical composition,
which are unrealistic in the case of helium dominated atmosphere. However, the
detailed study of this problem is beyond the scope of the present paper, which
is concerned 
with
the flux variations in individual passbands and not in the
differences between magnitudes of individual passbands.

\paragraph{Light absorption (extinction) in the circumstellar environment}

\citet{towog} were able to simulate successfully  the light curve of another
helium-rich chemically peculiar star \object{$\sigma$~Ori~E} \citep[see
also][]{nakaji,smigro} using the model of rigidly rotating magnetosphere
\citep{towo}. However, we argue that our model of a rotating spotted star is
more convenient for \hvezda. First, our method allows 
the simulation of
the light
curve just from the surface chemical distribution and we do not use any free
parameter to fit the amplitude. Second, the light curve of \hvezda\ is quite
different from that of $\sigma$~Ori~E, which shows deep and relatively narrow
minima with maximum amplitude in the $u$ \citep{peto}. Finally, $\sigma$~Ori~E
is known for its emission in H$\alpha$ line \citep{sigoriee}, which is absent in
the case of \hvezda. The uneven distribution of chemical elements on the surface
of $\sigma$~Ori~E \citep{prseni} may, however, 
also influence
the light curve of
$\sigma$~Ori~E and may explain some secondary effects which are not explained by
the cloud absorption model.

\paragraph{Light variations of other chemically peculiar stars}

Since silicon is overabundant in the majority of chemically peculiar stars
that show light variations (Si, He-weak, and He-strong stars, very frequently
cool CP stars, e.g., \citealt{jednadvacet}) it may play a crucial role
also in their variability. However, this has to be tested also for
other CP stars, as, e. g. Carpenter (1985) showed that the anomalous Zeeman
effect is capable to induce such an enhancement of the Si {\sc ii}
4131\AA~without the need of the large overabundance of the element.

\section{Conclusions}

In this work we were able to explain consistently the light curve of He-strong
chemically peculiar  star \hvezda. The observed light variations are explained
in terms of the uneven surface distribution of helium and silicon. The helium
and silicon spots modify the emergent flux predominantly due to the bound-free
transitions. The predicted light curves reproduce the observed ones very well in
their overall shape and amplitude. Although the agreement between the observed
and predicted light curves may seem satisfactory, we stress that there are also
other effects (e.g., NLTE effects, influence of other chemical elements) that
may be important for the light curve.

We have shown that bound-free transitions of mainly heavier elements (in our
case silicon) accompanied with uneven surface distribution of these elements are
crucial for the light variability of magnetic chemically peculiar stars.
To our knowledge, this mechanism has never been invoked in the literature
to predict the light variability.

Our results are important not only from the point of view of explanation of
light curves of chemically peculiar stars, but the comparison of predicted and
observed light curves represents an important check of model atmospheres and
their predictions.

\begin{acknowledgements}
The authors thank the anonymous referee for his/her valuable comments and
suggestions on the manuscript. This work was supported by grants GA\,\v{C}R
205/06/0217, VEGA 2/6036/6, and MVTS \v{C}RSR 10/15. This research has made use
of NASA's Astrophysics Data System, the SIMBAD database, operated at the CDS,
Strasbourg, France and On-line database of photometric observations of mCP stars
\citep{mikdat}.
\end{acknowledgements}

\newcommand\poprad{2004, IAU Symp. No. 224, The A-Star Puzzle, eds. J. Zverko, J.
\v{Z}i\v{z}\v{n}ovsk\'y, S.J. Adelman,  W.\,W. Weiss (Cambridge: Cambridge
University Press)}

\end{document}